\documentclass[journal,10pt,draftclsnofoot,onecolumn]{IEEEtran}

\usepackage{amsmath}
\usepackage{amssymb}
\usepackage{algorithmicx}
\usepackage{algpseudocode}
\usepackage{algorithm}
\usepackage{graphicx}

\newtheorem{definition}{Definition}
\newtheorem{theorem}{Theorem}
\newtheorem{proposition}{Proposition}
\newtheorem{lemma}{Lemma}
\newtheorem{corollary}{Corollary}
\ifCLASSINFOpdf
\else
\fi
\begin{document}
%
\title{Combinatorial Algorithms for Control of Biological Regulatory Networks}
%
%
%

\author{Andrew Clark,~\IEEEmembership{Member,~IEEE,}
        Phillip Lee,~\IEEEmembership{Member,~IEEE,}
        Basel Alomair,~\IEEEmembership{Senior Member,~IEEE,}
        Linda Bushnell,~\IEEEmembership{Fellow,~IEEE,}
        and~Radha Poovendran,~\IEEEmembership{Fellow,~IEEE}
\thanks{A. Clark is with the Department of Electrical and Computer Engineering, Worcester Polytechnic Institute, Worcester, MA, 01609 USA. Email: aclark@wpi.edu}%
\thanks{P. Lee, L. Bushnell, and R. Poovendran are with the Network Security Lab, Department of Electrical Engineering, University of Washington, Seattle, MA, 98195 USA. Email: \{leep3, lb2,rp3\}@uw.edu}
\thanks{B. Alomair is with the National Center for Cybersecurity Technology, King Abdulaziz City for Science and Technology, Riyadh, Saudi Arabia. Email: alomair@kacst.edu.sa}}

\maketitle

\begin{abstract}
Biological processes, including cell differentiation, organism development, and disease progression, can be interpreted as attractors (fixed points or limit cycles) of an underlying networked dynamical system. In this paper, we study the problem of computing a minimum-size subset of control nodes that can be used to steer a given biological network towards a desired attractor, when the networked system has Boolean dynamics. We first prove that this problem cannot be approximated to any nontrivial factor unless P=NP. We then formulate a sufficient condition and prove that the sufficient condition is equivalent to a target set selection problem, which can be solved using integer linear programming. Furthermore, we show that structural properties of biological networks can be exploited to reduce the computational complexity. We prove that when the network nodes have threshold dynamics and certain topological structures, such as block cactus topology and hierarchical organization, the input selection problem can be solved or approximated in polynomial time. For networks with nested canalyzing dynamics, we propose polynomial-time algorithms that are within a polylogarithmic bound of the global optimum. We validate our approach through numerical study on real-world gene regulatory networks.
\end{abstract}

\section{Introduction}
\label{sec:intro}
Biological processes, including gene expression and metabolism, are driven by complex interactions between basic building blocks. These interactions are often modeled as networked dynamical processes, in which nodes represent genes or proteins, links represent regulation of one component by another, and the node states describe the level of expression of each gene or protein. One widely-studied modeling approach is to assign a Boolean (on or off) state to each node, while describing the state of a node at each time step as a Boolean function of the neighbor states at the previous time step~\cite{alon2006introduction}. This approach provides biologically relevant insights as well as computational tractability.

The finite set of possible states implies that, in a deterministic network, the Boolean dynamics will eventually converge to a sequence of states that repeat infinitely, which is denoted as an \emph{attractor} of the network~\cite{kauffman1993origins}. An attractor could be a single state (a fixed point) or a cycle consisting of multiple states. It has been shown that attractors have biological interpretations in different contexts, for example as different types of differentiated stem cells \cite{huang2005cell}, states of disease progression \cite{huang2009cancer}, or stages of the cell cycle \cite{davidich2008boolean, li2004yeast}. 

Reprogramming a stem cell, or driving a cell from a diseased to a healthy state, can be interpreted as applying a control input to steer the network to a desired attractor~\cite{shmulevich2010probabilistic}. Control can be applied to a biological regulatory network by targeting a subset of genes to activate or repress, e.g., through drug therapies. This has been interpreted in the Boolean network framework as pinning a set of genes to a fixed state, corresponding to the desired attractor~\cite{macarthur2009systems}. The targeted genes then influence the dynamics of their neighbors, eventually steering the entire network state towards the desired attractor.  

This approach to control requires selecting a subset of genes that are sufficient to reach the desired attractor from an arbitrary, potentially pathological, initial state. In order to ensure minimal invasiveness and reduce cost, this set of genes should be as small as possible. There are, however, computational challenges associated with selecting a set of targeted genes. First, the number of such sets is exponential in the network size, making exhaustive search impractical. Second, verifying that any given set guarantees convergence to an attractor requires, in the worst case, evaluating convergence from an exponential number of possible initial states. Third, regulatory networks are noisy environments, creating uncertainty in the system model. As a result, existing algorithms for gene selection are either based on approximations from linear system theory (e.g., input selection for controllability)~\cite{wu2015network}, which do not capture the dynamical properties of the regulatory network, or are based on heuristics that inherently cannot provide guarantees on the minimality of the chosen set or the convergence to the desired attractor \cite{kim2013discovery}.

In this paper, we propose combinatorial algorithms for selecting a subset of genes to control in order to guarantee convergence to a desired attractor. Our approach is to formulate sufficient conditions for convergence to an attractor that we prove are equivalent to a target set selection (TSS) problem \cite{ackerman2010combinatorial}. While TSS is also computationally hard, we identify additional network structures that are common in biological networks and can be exploited to develop efficient approximation algorithms. We make the following specific contributions:
\begin{itemize}
\item We formulate the problem of selecting a minimum-size subset of nodes in order to guarantee convergence to a desired attractor. We prove a negative result, namely, that there is no approximation guarantee possible for this problem under arbitrary node dynamics unless $P=NP$.

\item We construct a sufficient condition for convergence to a desired attractor and prove that selecting a minimum-size set that satisfies this condition can be mapped to a TSS problem. 

\item We study the resulting TSS problem under two widely-occurring classes of regulatory networks, namely networks with threshold dynamics and hierarchical structure, and networks with nested canalyzing dynamics. For each type of network, we formulate polynomial-time algorithms that exploit the network structure to provide provable optimality bounds.

\item We generalize our approach to Boolean networks with probabilistic and asynchronous dynamics. We also show that our approach can be used to select input nodes in order to guarantee convergence to a cyclic attractor.

\item We evaluate our approach on several real-world biological network datasets as well as randomly generated topologies. We find that our proposed approach requires fewer input nodes to achieve a desired attractor compared to existing heursitics.
\end{itemize}

The paper is organized as follows. Section \ref{sec:related} presents related work. Section \ref{sec:model} presents the system model and definitions, as well as background on the TSS problem. Section \ref{sec:formulation} contains the problem formulation and our proposed gene selection algorithms. Section \ref{sec:probabilistic} generalizes our approach to probabilistic and asynchronous networks. Section \ref{sec:simulation} contains our numerical study. Section \ref{sec:conclusion} concludes the paper.



\section{Related Work}
\label{sec:related}
 Boolean networks were developed as a computationally tractable approximation to ODE models of biological processes \cite{lahdesmaki2003learning,chaves2005robustness, karlebach2008modelling, bornholdt2008boolean}. The concept of attractors was introduced by Waddington \cite{waddington1940organisers} and further investigated by Kauffman \cite{kauffman1993origins}. The biological relevance of attractors has been confirmed by studies including \cite{huang2005cell,huang2009cancer}.  Methodologies for inferring regulatory networks from gene expression data have been proposed, e.g., \cite{lahdesmaki2003learning}. Generalizations to probabilistic networks were introduced in \cite{shmulevich2002probabilistic}. These works, however, do not consider the problem of selecting a subset of genes to control a regulatory network.

Existing works have modeled therapeutic interventions, such as drug therapies, as inputs to the regulatory network, with the goal of informing possible new treatments \cite{kim2013discovery,aswani2009graph,kearney2016framework}. Approaches based on breaking cycles in the network topology were proposed in \cite{aswani2009graph}. Since cycles are very common (indeed, over-represented compared to random networks with similar degree distributions \cite{alon2006introduction}), these approaches may be overly conservative. Heuristics such as genetic algorithms \cite{kim2013discovery} have also been proposed, but do not provide any guarantees on the optimality of the chosen set or on whether convergence to a desired attractor is guaranteed from any initial state.

The problem of controlling a Boolean network is related to selecting input nodes to control a networked dynamical system. Existing works, however, typically consider a linear system with continuous state variables through methods such as controllability analysis \cite{wu2015network, liu2011controllability}, which are not applicable to nonlinear biological networks with a limited range of possible control signals.

The target set selection (TSS) problem was first identified in the social networking community \cite{ackerman2010combinatorial,ben2011treewidth,chiang2013some}, and is known to be NP-hard and difficult to approximate \cite{ben2011treewidth}. Recent efforts to develop approximation algorithms have focused on special cases of the network topology, such as trees, sparse graphs, and cliques \cite{ben2011treewidth,chiang2013some}. To the best of our knowledge, application of TSS to biological networks, as well as algorithms that exploit the structural properties of biological structures to reduce the complexity of TSS, have not been studied.

\section{Model and Background}
\label{sec:model}
This section presents the gene regulatory network model, followed by background on the target set selection problem.

\subsection{Regulatory Network Model}
\label{subsec:model}
A regulatory network is modeled as a graph $G=(V,E)$ with node set $V$ equal to the set of genes and $E$ denoting the set of edges. The number of vertices $|V| = n$. The network topology is assumed to be directed, with an edge $(i,j)$ implying that node $i$ regulates node $j$. The in-degree of node $i$, denoted $N_{in}(i) = |\{j : (j,i) \in E\}|$ is equal to the number of nodes that regulate $i$, while the out-degree, defined as $N_{out}(i) = |\{j : (i,j) \in E\}|$ is equal to the number of nodes that are regulated by $i$. The graph may contain edges between a node $i$ and itself. For any subset $A \subseteq V$, we let $G(A) = (A, E(A))$, where $E(A) = E \cap A \times A$, denote the subgraph induced $A$. 

As a preliminary, for a graph $G=(V,E)$, if $V = V_{1} \cup \cdots \cup V_{m}$ is a disjoint partition of the node set, then we define the \emph{graph contraction} $\overline{G}$ to be a graph with vertex set $\{1,\ldots,m\}$ and an edge from $i$ to $j$ if the exists an edge $(u,v)$ with $u \in V_{i}$ and $v \in V_{j}$. Note that this construction could result in multiple edges between the same nodes in $\overline{G}$.

Each node $i$ has a Boolean, discrete-time state variable $x_{i}(t) \in \{0,1\}$. Let $\mathbf{x}(t) \in \{0,1\}^{n}$ denote the vector of node states. The state dynamics of node $i$ are given by $$x_{i}(t+1) = f_{i}(\mathbf{x}(t)),$$ where $f_{i} : \{0,1\}^{n} \rightarrow \{0,1\}$ is a function that determines the state of node $i$ at time $(t+1)$ as a function of its neighbors' states. The function $f_{i}$ satisfies $f_{i}(\mathbf{x}) = f_{i}(\mathbf{x}^{\prime})$ whenever $\mathbf{x}_{j} = \mathbf{x}_{j}^{\prime}$ for all $j \in N_{in}(i)$. The dynamics are written in network form as $\mathbf{x}(t+1) = f(\mathbf{x}(t))$, where $f(\mathbf{x}) = (f_{1}(\mathbf{x}),\ldots,f_{n}(\mathbf{x}))$. We define the notations $\vee$, $\wedge$, and $\neg$ denote Boolean OR, AND, and NOT, respectively.

We now discuss two important special cases of the Boolean dynamics $f_{i}$. A node $i$ has \emph{threshold dynamics} if $f_{i}$ satisfies 
\begin{displaymath}
f_{i}(\mathbf{x}) = \left\{
\begin{array}{ll}
1, & \sum_{j \in N_{in}(i)}{a_{ij}x_{j}} \geq \tau_{i} \\
0, & \mbox{else}
\end{array}
\right.
\end{displaymath}
where $\tau_{i}$ is a real-valued threshold and $a_{ij}$ are real-valued coefficients. A positive value of $a_{ij}$ represents an excitatory link from $j$ to $i$, while a negative value represents an inhibitory link. 

A node $i$ has \emph{nested canalyzing} dynamics if $f_{i}$ is defined as follows. Let $d$ denote the in-degree of node $i$, and let $j_{1},\ldots,j_{d}$ be an ordering of $N_{in}(i)$. Let $a_{1},\ldots,a_{d},a_{d+1} \in \{0,1\}$ and $b_{1},\ldots,b_{d} \in \{0,1\}$. Then the nested canalyzing dynamics are defined by 
\begin{displaymath}
f_{i}(\mathbf{x}) = \left\{
\begin{array}{ll}
a_{l}, & \mbox{if } x_{j_{1}} \neq b_{1},\ldots, x_{j_{l-1}} \neq b_{l-1}, x_{jl} = b_{l} \\
a_{d+1}, & \mbox{else}
\end{array}
\right.
\end{displaymath}
In words, under nested canalyzing dynamics, there is a ranking of inputs to node $i$. If the top-ranked neighbor $j_{1}$ is in state $b_{1}$, then node $i$ moves to state $a_{1}$. Otherwise, if the neighbor $j_{2}$ is in state $b_{2}$, then $i$ moves to state $a_{2}$, and so on. If none of the conditions are met, then node $i$ reverts to a default state $a_{d+1}$. An \emph{attractor} of a Boolean network is defined as follows.

\begin{definition}
\label{def:attractor}
An attractor of length $r$ is a sequence of states $\mathbf{x}^{1},\ldots,\mathbf{x}^{r}$ such that $\mathbf{x}^{l} = f(\mathbf{x}^{l-1})$ for $l=2,\ldots,r$ and $\mathbf{x}^{1} = f(\mathbf{x}^{r})$.
\end{definition}

An attractor is a set of states that repeat in the Boolean network, so that any network that reaches one of the states in the attractor will remain in the attractor. It can be shown that, for any initial state $\mathbf{x}(0)$, there exists a finite time $T$ such that $\mathbf{x}(T)$ belongs to an attractor.

Attractors can be further classified as fixed points (where $r=1$), limit cycles (where $r$ is small relative to the network size), and chaotic (where $r$ is large).

Lastly, the effect of supplying inputs is described as follows. The set of input nodes is denoted $S$. For any node $i \in S$, there is a variable $s_{i} \in \{0,1\}$ such that $x_{i}(t) \equiv s_{i}$ for all $t$, i.e., each input node is pinned to a fixed state for all time $t$. 

\subsection{The Target Set Selection (TSS) Problem}
\label{subsec:target_background}
The target set selection problem is defined as follows. Let $G=(V,E)$ be a graph, and suppose that each node $v \in V$ is assigned a threshold $\tau(v)$. Let $S$ be a subset of nodes.

Initialize set $X[0] = S$. At step $k$, for $k=1,2,\ldots,$ define a set $Y[k]$ by $$Y[k] = \left\{v \notin X[k-1]: |N_{in}(v) \cap X[k-1]| \geq \tau(v)\right\}$$ and set $X[k] = X[k-1] \cup Y[k]$.Clearly, $X[k] \subseteq X[l]$ when $k < l$. 

 This process converges when $X[k] = X[k+1]$ for some $k$. Let $X^{\ast}$ be the set $X[k]$ at the iteration when convergence occurs. If $X^{\ast} = V$, then the set $S$ is denoted as a \emph{target set} of the graph. The target set selection problem is the problem of choosing a minimum-cardinality set for a given graph and set of thresholds. 
 
 \begin{proposition}[\cite{ben2011treewidth}]
 \label{prop:TSS_complexity}
 The TSS problem is NP-hard.
 \end{proposition}

Although the target set selection problem is NP-hard, tractable algorithms have been found for specific graphs such as complete graphs and graphs with bounded tree-width \cite{ben2011treewidth}.
\section{Input Selection Problem Formulation}
\label{sec:formulation}
This section first formulates the minimum gene selection problem. We analyze the complexity of the problem and then identify a sufficient condition based on target set selection. We provide algorithms for the special cases of threshold and nested canalyzing dynamics.

\subsection{Problem Formulation and Complexity}
\label{subsec:formulation}

Consider a gene regulatory network defined by a graph $G=(V,E)$ and a set of Boolean functions $\{f_{i} : i \in V\}$. Let $\mathbf{x}^{\ast} \in \{0,1\}^{n}$ denote a fixed-point attractor of the network, i.e., a state satisfying $f(\mathbf{x}^{\ast}) = \mathbf{x}^{\ast}$. The case where the attractor consists of multiple states will be considered in Section \ref{sec:probabilistic}. 

\begin{definition}
\label{def:convergence}
We say that a set of inputs $S$ guarantees convergence to the desired attractor $\mathbf{x}^{\ast}$ if setting $x_{i}(t) \equiv x_{i}^{\ast}$ for all $t$ implies that $\mathbf{x}(T) = \mathbf{x}^{\ast}$ for $T$ sufficiently large, for any initial state $\mathbf{x}(0)$ with $x_{i}(0) = x_{i}^{\ast}$ when $i \in S$.
\end{definition}

We let $\mathcal{C}$ denote the collection of input sets that guarantee convergence to $\mathbf{x}^{\ast}$. The minimum gene selection problem is then formulated as 
\begin{equation}
\label{eq:formulation}
\begin{array}{ll}
\mbox{minimize} & |S| \\
\mbox{s.t.} & S \in \mathcal{C}
\end{array}
\end{equation}

We first analyze the complexity of the problem, and find that non-trivial approximations are impossible unless $P=NP$.

\begin{theorem}
\label{theorem:complexity}
If there exists a function $\gamma: \mathbb{N} \rightarrow \mathbb{N}$ and a polynomial-time algorithm that takes as input an instance of (\ref{eq:formulation}) and is guaranteed to output a set $S$ satisfying $S \in \mathcal{C}$ and $|S| \leq \gamma(n) |S^{\ast}|$, where $|S^{\ast}|$ is the optimal solution to (\ref{eq:formulation}), then $P=NP$.
\end{theorem}

\begin{IEEEproof}
The proof is by showing that, if there exists an algorithm that satisfies the conditions of the theorem, then that algorithm can also be used to solve 3-SAT, which is an NP-hard problem. An instance of the 3-SAT problem consists of determining whether, given a set of Boolean variables $\{q_{1},\ldots,q_{m}\}$, there exists a set of values for the $q_{i}$'s such that the relation
\begin{equation}
\label{eq:sat}
(p_{11} \vee \cdots \vee p_{1r_{1}}) \wedge \cdots \wedge (p_{l1} \vee \cdots \vee p_{lr_{l}}),
\end{equation}
 where $p_{ij} \in \{q_{s}, \neg q_{s} : s \in \{1,\ldots,m\}\}$ for all $i,j$, evaluates to true.
 

Suppose that an instance of 3-SAT is given. Construct a Boolean network as follows. Let $V = V_{1} \cup V_{2} \cup V_{3}$, where $V_{1}$ is indexed $v_{1}^{1},\ldots,v_{m}^{1}$, $V_{2}$ is indexed $v_{1}^{2},\ldots,v_{l}^{2}$, and $V_{3}$ is a singleton node $v^{3}$. 

The edge set is defined as follows. We add an edge $(v_{i}^{1},v_{j}^{2})$ if $p_{js} \in \{q_{i}, \neg q_{i}\}$ for some $s$. We add edges $(v_{i}^{1},v_{j}^{1})$ between all nodes in $V_{1}$. We include an edge $(v_{i}^{2},v^{3})$ for $i=1,\ldots,l$. Finally, we add an edge from $v^{3}$ to each other node.

For all nodes $v \in V$, we set $x_{v}(t+1) = 1$ if $x_{v^{3}}(t) = 1$. Otherwise, the dynamics are defined as follows. We choose the functions $f_{v_{i}^{1}}$ for $i=1,\ldots,m$  so that the binary string $x_{v_{1}^{1}}(t)x_{v_{2}^{1}}(t)\cdots x_{v_{m}^{1}}(t)$ satisfies 
\begin{displaymath}
x_{v_{1}^{1}}(t+1)x_{v_{2}^{1}}(t+1)\cdots x_{v_{m}^{1}}(t+1) 
= x_{v_{1}^{1}}(t)\cdots x_{v_{m}^{1}}(t) + 1 \pmod{2^{m}}.
\end{displaymath}
 We set $$x_{v_{i}^{2}}(t+1) = \bigvee_{s=1}^{r_{i}}{y_{is}(t)},$$ where $y_{is}(t) = x_{j}(t)$ if $p_{is} = q_{j}$ and $y_{is}(t) = \neg x_{j}(t)$ if $p_{is} = \neg q_{j}$. Furthermore, we set $x_{v^{3}}(t+1) = 1$ if $x_{v_{i}^{2}}(t)=1$ for all $i$.



Suppose first that there is a solution to (\ref{eq:sat}), equal to $q_{1} = \overline{q}_{1},\ldots, q_{m} = \overline{q}_{m}$. By construction of the network, for any initial state, eventually there will exist time $T$ such that $x_{v_{i}^{1}}(T) = \overline{q}_{i}$ for all $i=1,\ldots,m$, and hence $x_{v^{3}}(T+2) = 1$ and $x_{v}(t) = 1$ for all $v \in V$ and $t \geq T+3$. Hence, if there is a solution to (\ref{eq:sat}), then $S^{\ast} = \emptyset$.

On the other hand, suppose there is no solution to (\ref{eq:sat}), and consider an initial state with $x_{v_{i}^{2}}(0) = 0$ for all $i$, $x_{v^{3}}(0) = 0$, and all other initial states arbitrary. Since there is no solution to (\ref{eq:sat}), $x_{v_{i}^{2}}(t) = x_{v^{3}}(t) = 0$ for all $t$, implying that the attractor is never reached when $S = \emptyset$. Convergence to the desired attractor can, however, be achieved by setting $S = \{v^{3}\}$. Hence $S^{\ast} = \{v^{3}\}$ is a minimum-size solution to (\ref{eq:formulation}) and $S^{\ast} = \emptyset$ iff the relation (\ref{eq:sat}) is not satisfiable. 

If there is a deterministic polynomial-time algorithm that returns a set $S$ satisfying $|S| \leq \gamma(n) |S^{\ast}|$ for some $\gamma$, then that algorithm must choose $S = \emptyset$ whenever $S^{\ast} = \emptyset$. Conversely, $S^{\ast} \neq \emptyset$, then $S \neq \emptyset$. Hence, we can construct a polynomial-time algorithm for 3-SAT by following the above procedure to construct a gene regulatory network, and outputting true if the algorithm returns $\emptyset$ and false otherwise. Thus, if there exists such an algorithm, then $P=NP$.
\end{IEEEproof}

The proof of Theorem \ref{theorem:complexity} also implies that it is NP-hard to verify whether a given set of input nodes $S$ guarantees convergence to a desired attractor. If not, then it would be possible to check whether there exists a solution to 3-SAT (an NP-complete problem) by verifying whether $S = \emptyset$ guarantees convergence to $\mathbf{x}^{\ast} = \mathbf{1}$ in the graph constructed above.

\subsection{Mapping to Target Set Selection}
\label{subsec:TSS_mapping}

In order to develop efficient approximation algorithms for relevant special cases of (\ref{eq:formulation}), we first introduce a sufficient condition that is equivalent to a target set selection problem.

Given a gene regulatory network $G=(V,E)$ and dynamics $f_{i}$ for $i=1,\ldots,n$, we construct an extended network $\hat{G} = (\hat{V}, \hat{E})$ as  follows. For each Boolean function $f_{i}$, we can write $f_{i}$ in conjunctive normal form  as $$f_{i}(\mathbf{x}) = (y_{11} \vee \cdots \vee y_{1r_{1i}}) \wedge \cdots \wedge (y_{l_{i}1} \vee \cdots \vee y_{l_{i}r_{li}}),$$ where $y_{is} \in \{x_{j}, \neg x_{j}\}$ for some $j \in N(i)$. The node set $\hat{V}$ is defined by $$\hat{V} = V \cup \{a_{i_{s}} : s = 1,\ldots,l\}.$$ The edge set $\hat{E}$ is defined by 
\begin{IEEEeqnarray*}{rCl}
\hat{E} &=& \{(a_{i_{s}},i) : s=1,\ldots,l\} \cup \{(j, a_{i_{s}}) : x_{j} \in \{y_{su} : u=1,\ldots,l_{s}\}, x_{j}^{\ast} = x_{i}^{\ast}\} \\
&&  \cup \{(j, a_{i_{s}}) : \neg x_{j} \in \{y_{su} : u=1,\ldots,l_{s}\}, x_{j}^{\ast} \neq x_{i}^{\ast}\}.
\end{IEEEeqnarray*}


In words, we have an edge from each $a_{i_{s}}$ to $i$. We also have an edge from $j$ to $a_{i_{s}}$ if $j$ inhibits $i$ and $x_{i}^{\ast} \neq x_{j}^{\ast}$, and an edge from $j$ to $a_{i_{s}}$ if $j$ activates $i$ and $x_{i}^{\ast} = x_{j}^{\ast}$.

The thresholds for this augmented graph are given by
\begin{displaymath}
\tau(a_{i_{m}}) = \left\{
\begin{array}{ll}
r_{mi}, & x_{i}^{\ast} = 0 \\
1, & x_{i}^{\ast} = 1
\end{array}
\right. \quad \tau(i) = \left\{
\begin{array}{ll}
1, & x_{i}^{\ast} = 0 \\
l_{i}, & x_{i}^{\ast} = 1
\end{array}
\right.
\end{displaymath}

The threshold that is chosen depends on whether each node is on or off in the desired attractor. 
We now prove that solving   target set selection on this augmented graph is sufficient for ensuring convergence to a desired attractor.

\begin{proposition}
\label{prop:TSS_sufficient}
Suppose that $S \subseteq V$ is a solution to the target set selection problem on the augmented graph $\hat{G}$. Then  $S \in \mathcal{C}$.
\end{proposition}

The proof is omitted due to space constraints. We observe that this condition also gives a polynomial-time algorithm for checking whether a given set of inputs guarantees convergence, namely, allowing the target set dynamics $X[k]$ to unfold for $2n$ iterations on the graph $\hat{G}$. While $V \subseteq X[n]$ implies convergence to the desired attractor from any initial state, the converse is not necessarily true.

We now compare this sufficient condition to a known sufficient condition from previous work \cite{aswani2009graph}.

\begin{proposition}
\label{prop:weaker}
Suppose that a set $S$ satisfies $S \cap T \neq \emptyset$ for all cycles $T$ in the graph $G$ and each node is path-connected in $G$ to at least one node in $S$. Then the set $S$ is a target set for the augmented graph $\hat{G}$.
\end{proposition}

\begin{IEEEproof}
Suppose that $S$ satisfies the conditions of the theorem, and yet $S$ is not a target set. Then there exists $i \in V$ such that $i \notin X[k]$ for any $k$. We must have $i \notin S$. If $i$ has no neighbors in $\hat{V}$, then $i$ is not connected to any input, a contradiction. Otherwise, by construction, there exists at least one $a_{i_{s}}$ such that $a_{i_{s}} \notin X[k]$, and therefore at least one neighbor $j \in N(i)$ such that $x_{j} \notin X^{\ast}$. If $j = i$, then there is a cycle $T$, consisting of the self-loop $(i,i)$, such that $S \cap T = \emptyset$, a contradiction. 

Proceeding inductively, we maintain a set $U \subseteq V \setminus X^{\ast}$ where each node in $U$ is path-connected to $i$. Within $2n$ iterations, we must have a node that is added to $U$ twice, implying the existence of a cycle in the graph that is disjoint from $S$ and yielding a contradiction.
\end{IEEEproof}

Proposition \ref{prop:weaker} implies that the target set selection condition is weaker (easier to satisfy) than the current known approach of selecting a subset of nodes to break all cycles in the graph. On the other hand, this approach also requires adding new nodes and edges to the underlying the graph, and moreover, relies on solving the computationally difficult TSS problem. In the general case, this problem can be formulated as the integer linear program \cite{ackerman2010combinatorial}

\begin{equation}
\label{eq:TSS_IP}
\begin{array}{ll}
\mbox{minimize} & \sum_{i=1}^{n}{s_{i}} \\
\mathbf{s}, \mathbf{e} & \\
\mbox{s.t.} & \sum_{(i,j) \in \hat{E}}{e_{ij}} \geq \overline{\tau}_{i}(1-s_{i}) \ \forall i \in \hat{V} \\
 & e_{ij} + e_{ji} = 1 \ \forall i \neq j \\
 & e_{ij} + e_{jl} + e_{li} \leq 2 \ \forall i,j,l \mbox{ distinct} \\
 & e_{ij} \in \{0,1\}, s_{i} \in \{0,1\}
\end{array}
\end{equation}

The binary variables $\{s_{i}: i=1,\ldots,n\}$ satisfy $s_{i} =1$ iff $i \in S$. Hence, the solution to TSS can be obtained by solving (\ref{eq:TSS_IP}) and selecting the set $S$ based on the $s_{i}$'s.

In the next subsections, we examine biologically relevant special cases and develop algorithms that exploit these additional structures. 




 
 
 
 
 \subsection{Threshold Dynamics}
 \label{subsec:threshold}
 We now analyze the target set selection formulation in the special case where the nodes have threshold dynamics. Specifically, we assume threshold dynamics where, for each node $i$, either $a_{ij} = 1$ for all $j \in N_{out}(i)$ or $a_{ij} = -1$ for all $j \in N_{out}(i)$. Intuitively, all nodes are either purely excitatory ($a_{ij}=1)$ or purely inhibitory ($a_{ij}=-1$), and all nodes exert the same effect on their neighbors. Let $\mathcal{E} \triangleq \{i: a_{ij} = 1 \ \forall j \in N_{out}(i)\}$ and $\mathcal{I} \triangleq \{i : a_{ij} = -1 \ \forall j \in N_{out}(i)\}$.  For the given attractor $\mathbf{x}^{\ast}$, define $\mathcal{E}^{1} \triangleq \mathcal{E} \cap \{i : x_{i}^{\ast} = 1\}$ and $\mathcal{E}^{0} \triangleq \mathcal{E} \cap \{i : x_{i}^{\ast} = 0\}$, and define $\mathcal{I}^{1}$ and $\mathcal{I}^{0}$ in an analogous manner. 
 
 In the case of threshold dynamics, the target set selection instance is as follows. The graph $\hat{G}$ has node set $\hat{V} = V$. An edge $(i,j)$ exists from node $i$ to node $j$ if $x_{i}^{\ast} = x_{j}^{\ast}$ and node $i \in \mathcal{E}$, or if $x_{i}^{\ast} \neq x_{j}^{\ast}$ and $i \in \mathcal{I}$. The threshold $\hat{\tau}(i)$ is given by
 \begin{displaymath}
 \hat{\tau}(i) = \left\{
 \begin{array}{ll}
 \tau(i) + |N_{in}(i) \cap \mathcal{I}|, & x_{i}^{\ast} = 1 \\
 \tau(i) + |N_{in}(i) \cap \mathcal{E}|, & x_{i}^{\ast} = 0
 \end{array}
\right.
 \end{displaymath}

 
 \begin{lemma}
 \label{lemma:TSS_threshold}
 Suppose that $S$ is a target set for the graph $\hat{G}$ with thresholds $\hat{\tau}$. Then there exists $T$ such that $\mathbf{x}(t) = \mathbf{x}^{\ast}$ for all $t \geq T$.
 \end{lemma}

\begin{IEEEproof}
We show that if $i \in X[k]$ for some $k > 0$, then $x_{i}(t)$ converges to $x_{i}^{\ast}$. The proof is by induction on $k$, noting that $X[0] = S$. At time $k$, suppose that $i \in X[k] \setminus X[k-1]$, and hence the threshold condition is satisfied. We have that $$|\hat{N}_{in}(i) \cap X[k-1]| \geq \tau(i) + |N_{in}(i) \cap \mathcal{I}|,$$ which is equivalent to
\begin{multline}
\label{eq:TSS_threshold}
|\hat{N}_{in}(i) \cap \mathcal{E}^{1} \cap X[k-1]| + |\hat{N}_{in}(i) \cap \mathcal{E}^{0} \cap X[k-1]| \\
 + |\hat{N}_{in}(i) \cap \mathcal{I}^{1} \cap X[k-1]| + |\hat{N}_{in}(i) \cap \mathcal{I}^{0} \cap X[k-1]|
  \geq \tau(i) + |N_{in}(i) \cap \mathcal{I}|.
\end{multline}
Suppose that $x_{i}^{\ast} = 1$; the case where $x_{i}^{\ast} = 0$ is similar. Then (\ref{eq:TSS_threshold}) is equivalent to 
\begin{displaymath}
|N_{in}(i) \cap \mathcal{E}^{1} \cap X[k-1]| + |N_{in}(i) \cap \mathcal{I}^{0} \cap X[k-1]| 
\geq \tau(i) + |N_{in}(i) \cap \mathcal{I}|.
\end{displaymath}
 By inductive hypothesis, for $t$ sufficiently large, $x_{j}(t) = x_{j}^{\ast}$ for all $j \in X[k-1]$. Thus for $t$ sufficiently large, we have 
\begin{eqnarray*}
\sum_{j \in N_{in}(i)}{a_{ij}x_{j}(t)} &=& \sum_{j \in N_{in}(i) \cap \mathcal{E}}{x_{j}(t)} - \sum_{j \in N_{in}(i) \cap \mathcal{I}}{x_{j}(t)} \\
&=& \sum_{j \in N_{in}(i) \cap \mathcal{E}}{x_{j}(t)}  + \sum_{j \in N_{in}(i) \cap \mathcal{I}}{(1-x_{j}(t)} - |N_{in}(i) \cap \mathcal{I}| \\
&\geq& |N_{in}(i) \cap \mathcal{E}^{1} \cap X[k-1]| 
 + |N_{in}(i) \cap \mathcal{I}^{0} \cap X[k-1]| - |N_{in}(i) \cap \mathcal{I}| \\
&\geq& \tau(i) + |N_{in}(i) \cap \mathcal{I}| - |N_{in}(i) \cap \mathcal{I}| = \tau(i).
\end{eqnarray*}
Hence $x_{i}(t) = 1$ for $t$ sufficiently large, implying that the desired attractor is reached. The fact that $X^{\ast} = V$ completes the proof.
\end{IEEEproof}

Under threshold dynamics, formulating the input selection problem as TSS does not require adding any nodes to the graph $G$. 
We analyze algorithms for computing optimal target sets under network topologies that typically arise in biological networks. We first consider complete graphs, which arise as subgraphs of regulatory networks, and consider a generalization of known results to networks with both positive and negative edges.

If the graph $G$ is complete, then the edge set will be given by 
\begin{displaymath}
\hat{E} = \{(i,j) : i \in \mathcal{E}^{1} \cup \mathcal{I}^{0}, j \in \mathcal{E}^{1} \cup \mathcal{I}^{1}\} \\
\cup \{(i,j) : i \in \mathcal{E}^{0} \cup \mathcal{I}^{1}, j \in \mathcal{E}^{0} \cup \mathcal{I}^{1}\}
\end{displaymath}



\begin{lemma}
\label{lemma:clique_helper}
Let $S$ be a minimum-size target set for a complete graph. Suppose that there exist two nodes $i$, $j$ such that $i$ and $j$ both lie in $\mathcal{E}^{1}$, $\mathcal{E}^{0}$, $\mathcal{I}^{1}$, or $\mathcal{I}^{0}$, $i \in S$, $j \notin S$, and $\hat{\tau}(i) < \hat{\tau}(j)$. Then $S \setminus \{i\} \cup \{j\}$ is also a solution to the target set selection problem.
\end{lemma}

\begin{IEEEproof}
Let $X[k]$ denote the thresholding process on $\hat{G}$ when the initial set  $X[0] = S$, and let $\overline{X}[k]$ denote the thresholding process when the initial set  $\overline{X}[0] = \overline{S} = S \setminus \{i\} \cup \{j\}$. We will show that the result holds when $\{i,j\} \subseteq \mathcal{E}^{1}$; other cases are similar and omitted due to space constraints. 

It suffices to show that $i \in \overline{X}[k]$ for some $k$. This will hold when 
\begin{equation}
\label{eq:clique_helper1}
|\mathcal{E}^{1} \cap \overline{X}[k-1]| + |\mathcal{I}^{0} \cap \overline{X}[k-1]| \geq \overline{\tau}(i).
\end{equation}
 Define 
\begin{eqnarray*}
\overline{\alpha}_{k} &=& |\mathcal{E}^{1} \cap \overline{X}[k-1]| + |\mathcal{I}^{0} \cap \overline{X}[k-1]| \\
\alpha_{k} &=& |\mathcal{E}^{1} \cap X[k-1]| + |\mathcal{I}^{0} \cap X[k-1]| \\
\overline{\beta}_{k} &=& |\mathcal{E}^{0} \cap \overline{X}[k-1]| + |\mathcal{I}^{1} \cap \overline{X}[k-1]|\\
\beta_{k} &=& |\mathcal{E}^{0} \cap X[k-1]| + |\mathcal{I}^{1} \cap X[k-1]|
\end{eqnarray*}
Now, since it is assumed that $S$ is a target set and $j \notin S$, we must have $\alpha_{k} \geq \overline{\tau}(j)$ for some $k$. Hence, if we can show that $\overline{\alpha}_{k} \geq \alpha_{k}$ and $\overline{\beta}_{k} \geq \beta_{k}$ for all $k$, then (\ref{eq:clique_helper1}) will be satisfied, since $\overline{\alpha}_{k} \geq \alpha_{k} \geq \overline{\tau}(j) \geq \overline{\tau}(i)$ for some $k$. 

The proof that $\overline{\alpha}_{k} \geq \alpha_{k}$ and $\overline{\beta}_{k} \geq \beta_{k}$ is by induction on $k$. When $k=1$, $X[0] = S$ and $\overline{X}[0] = \overline{S}$, and hence $\alpha_{k} = \overline{\alpha}_{k}$ and $\overline{\beta}_{k} = \beta_{k}$ by definition of $S$ and $\overline{S}$.  For larger values of $k$, we have that $$\mathcal{E}^{1} \cap \overline{X}[k-1] = \{s \in \mathcal{E}^{1} : \overline{\tau}(s) < \overline{\alpha}_{k-1}\} \cup (S \cap \mathcal{E}^{1}),$$ with similar identities for $\mathcal{E}^{0}$, $\mathcal{I}^{1}$, and $\mathcal{I}^{0}$. If $\overline{\alpha}_{k-1} \geq \alpha_{k-1}$ and $\overline{\beta}_{k-1} \geq \beta_{k-1}$, then $(\mathcal{I}^{0} \cap X[k])  \subseteq (\mathcal{I}^{0} \cap \overline{X}[k]$, with similar identities for $\mathcal{E}^{0}$ and $\mathcal{I}^{1}$.

To show that $(\mathcal{E}^{1} \cap X[k]) \subseteq (\mathcal{E}^{1} \cap \overline{X}[k])$, we first have that $$\{s \in \mathcal{E}^{1} :  \overline{\tau}(s) < \alpha_{k-1}\} \subseteq \{s \in \mathcal{E}^{1} :  \overline{\tau}(s) < \overline{\alpha}_{k-1}\}.$$ Hence $((\mathcal{E}^{1} \cap X[k-1]) \setminus (\mathcal{E}^{1} \cap \overline{X}[k-1])) \subseteq \{i\}.$ If the above holds with equality, then $\overline{\alpha}_{k-1} < \overline{\tau}(i)$, and hence by inductive hypothesis $\alpha_{k-1} < \overline{\tau}(j)$, implying that $j \notin (\mathcal{E}^{1} \cap X[k-1]$. We therefore have that 
\begin{displaymath}
|((\mathcal{E}^{1} \cap X[k-1]) \setminus (\mathcal{E}^{1} \cap \overline{X}[k-1]))| 
\leq |((\mathcal{E}^{1} \cap \overline{X}[k-1]) \setminus (\mathcal{E}^{1} \cap X[k-1]))|
\end{displaymath}
and hence $(\mathcal{E}^{1} \cap X[k]) \subseteq (\mathcal{E}^{1} \cap \overline{X}[k])$. Thus $\overline{\alpha}_{k} \geq \alpha_{k}$ and $\overline{\beta}_{k} \geq \beta_{k}$ for all $k$, completing the proof.
\end{IEEEproof}

Lemma \ref{lemma:clique_helper} implies that, when the graph is a clique, we can restrict the search space by ordering the vertices in $\mathcal{E}^{1}$, $\mathcal{E}^{0}$, $\mathcal{I}^{0}$ and $\mathcal{I}^{1}$ based on their thresholds, and choosing the $m(\mathcal{E}^{1})$ (resp. $m(\mathcal{E}^{0})$, $m(\mathcal{I}^{1})$, $m(\mathcal{I}^{0})$) vertices with largest threshold from $\mathcal{E}^{1}$ (resp. $\mathcal{E}^{0}$, $\mathcal{I}^{1}$, $\mathcal{I}^{0}$). For a network of $n$ nodes, there are no more than $n^{4}$ sets of this type, implying that the selection problem on a clique can be solved in $O(n^{4})$ time.

Next, we consider graphs that have a block cactus structure, in which the set of vertices $V$ can be partitioned as $V = V_{1} \cup \cdots \cup V_{m}$, where the subgraph induced by each $V_{i}$ is a clique, and the contraction of the graph around the $V_{i}$'s is a tree. 

\begin{proposition}
\label{prop:tree-of-cliques}
There exists an $O(n^{4})$ algorithm for computing a minimum-size set of input genes in a block cactus graph.
\end{proposition}

\begin{IEEEproof}
The proof is by induction on $m$. When $m=1$, the problem reduces to selecting a minimum-size input set for a clique. Suppose that the result holds up to $(m-1)$, and consider a tree of $m$ cliques. Without loss of generality, suppose that $V_{m}$ corresponds to a leaf in the tree, i.e., there is exactly one incoming edge that is not part of the clique. This assumption is without loss of generality because at least one node in the tree must be a leaf (a degree-one vertex), and hence we can always have $V_{m}$ as a leaf by reordering vertices.

We consider two cases on $V_{m}$. First, suppose that the only edge incident on $V_{m}$ is an incoming edge onto a node denoted $v \in V_{m}$. Define a new threshold vector for $V_{m}$ by $\hat{\tau}(v) = \overline{\tau}(v)-1$ and $\hat{\tau}(u) = \tau(u)$ for $u \neq v$. Let $S_{1}$ be a minimum-size input set for $V_{1} \cup \cdots \cup V_{m-1}$, and let $S_{2}$ be a minimum-size input set for $V_{m}$ when the threshold is equal to $\hat{\tau}(v)$. We then have that $S_{1} \cup S_{2}$ is a minimum-size input set for $G$, and the computation time is equal to $O((n-|V_{m}|)^{4}) + O(|V_{m}|^{4}) = O(n^{4})$. 

Conversely, suppose that the only edge incident on $V_{m}$ is an outgoing edge. Define a new threshold vector for $V_{1} \cup \cdots \cup V_{m-1}$ as $\overline{\tau}(v) = \tau{\tau}(v)-1$, where $v$ is the node that has an incoming edge from $V_{m}$. Then the selection algorithm is equivalent to choosing a set of nodes to ensure that $V_{m}$ reaches the desired attractor and a set of nodes to ensure that $V_{1} \cup \cdots \cup V_{m-1}$ reaches the desired attractor with thresholds $\hat{\tau}$, requiring only $O(n^{4})$ time in total.
\end{IEEEproof}

The algorithm for selecting a minimum-size set of input nodes  is described as Algorithm \ref{algo:threshold-tree}.

  \begin{center}
\begin{algorithm}[!htp]
	\caption{Algorithm for selecting a minimum-size set of genes to control a network with block cactus structure and threshold dynamics.}
	\label{algo:threshold-tree}
	\begin{algorithmic}[1]
		\Procedure{Threshold\_Selection}{$G$, $V_{1},\ldots,V_{m}$, $\overline{\tau}$}
            \State \textbf{Input}: Graph topology $G=(V,E)$, partition into cliques $V_{1},\ldots,V_{m}$, threshold vector $\overline{\tau}$
            \State \textbf{Output}: Minimum-size input set $S$
            \State \textbf{Assumption}: $V_{m}$ is a leaf in the tree
            \State $S_{2} \leftarrow $ minimum size set to control $V_{m}$.
            \If{$m==1$}
            \State \Return{$S_{2}$}
            \EndIf
            \State $v \leftarrow$ vertex of $V \setminus V_{m}$ with incoming edge from $V_{m}$.
            \State $\hat{\tau} \leftarrow \overline{\tau}$ restricted to $V_{1} \cup \cdots \cup V_{m}$
            \State $\hat{\tau}(v) \leftarrow \overline{\tau}(v) -1$
            \State $S_{1} \leftarrow$ Threshold\_Selection($G(V_{1} \cup \cdots \cup V_{m-1})$, $V_{1},\ldots,V_{m-1}$, $\hat{\tau}$)
            \State $S \leftarrow S_{1} \cup S_{2}$
            \State \Return{$S$} 
           	\EndProcedure
	 \end{algorithmic}
\end{algorithm}
\end{center}





Networks that do not have block cactus structure can be addressed within this framework by grouping the network nodes into densely-connected clusters, denoted $V_{1},\ldots, V_{m}$ (e.g., via the methods in \cite{herrero2001hierarchical}). For any two nodes $i$ and $j$ in the same cluster that are not connected by an edge, add an edge and increment $\hat{\tau}(i)$ and $\hat{\tau}(j)$ by $1$. Then, find a set of edges $E^{\prime}$ to remove in order to remove cycles from the contracted graph; such a set of edges corresponds to a minimum arc feedback set. The thresholds are unchanged at this stage.


In addition to consisting of loosely connected components, biological networks also often have modular, hierarchical structure. These modular structures are believed to be derived from the functional organization of cells. In the following, we analyze gene selection algorithms on a model of hierarchical networks introduced in \cite{ravasz2002hierarchical}. We first define the model as follows.

The hierarchical network is constructed iteratively. The network originates with a single hub node. At the first iteration, $k$ nodes are added and are connected to each other, creating a network $G_{1}$. At the $i$-th iteration, $k$ copies of $G_{i-1}$ are generated and connected to the hub node. An algorithm for constructing a minimum-size set of genes to control a hierarchical network. The graph is undirected, implying that for each edge $(i,j)$, there is a corresponding edge $(j,i)$.

  \begin{center}
\begin{algorithm}[!htp]
	\caption{Algorithm for selecting a minimum-size set of genes to control a network with hierarchical structure.}
	\label{algo:threshold-hierarchy}
	\begin{algorithmic}[1]
		\Procedure{Threshold\_Hierarchy}{$G=(V,E)$, $\overline{\tau}$}
            \State \textbf{Input}: Graph topology $G=(V,E)$, threshold vector $\overline{\tau}$
            \State \textbf{Output}: Minimum-size input set $S$
           \State $v \leftarrow$ hub of graph $G$
           \State $d \leftarrow$ depth of hierarchical network
           \State $G^{1},\ldots,G^{k} \leftarrow$ copies of network at depth $(d-1)$ 
           \State $S \leftarrow \emptyset$, $\Gamma \leftarrow \{1,\ldots,m\}$
           \For{$i=1,\ldots,k$}
           \State $\overline{\tau}^{i} \leftarrow$ threshold vector for graph $G_{i}$ 
           \State $\underline{\tau}^{i} \leftarrow \overline{\tau}^{i}-1$
           \State $\underline{S}_{i} \leftarrow Threshold\_Hierarchy(G_{i}, \underline{\tau}^{i})$
           \State $\overline{S}_{i} \leftarrow Threshold\_Hierarchy(G_{i}, \overline{\tau}^{i})$
           \State $c_{i} \leftarrow |\overline{S}_{i}| - |\underline{S}_{i}|$
           \If{$c_{i}=0$}
               \State $S \leftarrow S \cup \overline{S}_{i}$, $\Gamma \leftarrow \Gamma \setminus \{i\}$
           \EndIf
           \EndFor
           \If{$v$ not activated by target set $S$}
           \State $S \leftarrow S \cup \{v\}$
           \EndIf
           \For{$i \in \Gamma$}
           \State $S \leftarrow S \cup \underline{S}_{i}$
           \EndFor
           \State \Return{$S$}
           	\EndProcedure
	 \end{algorithmic}
\end{algorithm}
\end{center}

The algorithm is recursive. For a hierarchical network with $d$ levels, each consisting of $m$ copies, the approach is to compute, for each copy, the number of input nodes required with and without the central hub node in $S$. For all the sub-graphs where the number of input nodes is the same under both cases, select a subset of input nodes to drive the subgraph to the desired state. If the nodes in these sub-graphs are insufficient to drive the hub node over its threshold, then add the central hub node to the input set, recompute all remaining thresholds, and compute sets of inputs to guarantee that the remaining subgraphs reach the desired state.



\begin{proposition}
\label{prop:hierarchical_optimal}
Algorithm \ref{algo:threshold-hierarchy} selects a set $S$ satisfying $|S| \leq |S^{\ast}|\log{n}$, where $S^{\ast}$ is the minimum-size input set, in $O(n^{2})$ time.
\end{proposition}

\begin{IEEEproof}
To show the complexity, let $R(k,m)$ denote the number of computations required to compute the optimal set in a graph with $k$ copies and $m$ iterations. We have that $R(k,m+1) = 2k R(k,m)$, and hence $R(k,m) = (2k)^{m}$. At the same time, the number of nodes is equal to $n=(k+1)^{m}$. Hence we have $$\frac{R(k,m)}{n} = \frac{(2k)^{m}}{(k+1)^{m}} \leq 2^{m} = n,$$ and hence $R(k,m) = n^{2}$. 

We then analyze the optimality gap. Let $\epsilon(m)$ denote the worst-case optimality gap in a network with $m$ levels of hierarchy. We then have that $$|S| \leq \sum_{i=1}^{l}{|S \cap V_{i}|} + 1 \leq \sum_{i=1}^{m}{\epsilon(m)|S_{i}^{\ast}|} + 1.$$ On the other hand, $|S^{\ast}| \geq \sum_{i=1}^{m}{|S_{i}^{\ast}|}$, and hence combining these expressions yields 
\begin{eqnarray*}
\frac{|S|}{|S^{\ast}|} &\leq& \frac{\sum_{i=1}^{m}{\epsilon(m)|S_{i}^{\ast}|} + 1}{\sum_{i=1}^{m}{|S_{i}^{\ast}|}} \\
&=& \epsilon(m) + \frac{1}{\sum_{i}{|S_{i}^{\ast}|}} \leq \epsilon(m)+1,
\end{eqnarray*}
implying that $\epsilon(m+1) \leq \epsilon(m)+1$ and hence $\epsilon(m) \leq m$. Since $m = \frac{\log{n}}{\log{(k+1)}} \leq \log{n}$, we have the desired optimality bound.
\end{IEEEproof}

\subsection{Nested Canalyzing Dynamics}
\label{subsec:NC}
We now present sufficient conditions for selecting genes to ensure convergence in networks with nested canalyzing dynamics, as defined in Section \ref{subsec:model}. We first characterize the sufficient condition of Proposition \ref{prop:TSS_sufficient} for this class of dynamics.

\begin{lemma}
\label{lemma:NC_form}
For each node $i$, define $\Omega_{i} = \{r: a_{r} = x_{i}^{\ast}\}$, with $r_{i} = |\Omega_{i}|$. Then the following instance of TSS is sufficient to ensure convergence to the desired attractor. For each node $i$, define a collection of nodes $u_{i,1},\ldots,u_{i,r_{i}}$. Each node $u_{i,a_{s}}$ has an incoming edge from each node $j_{l} \in N(i)$ with $l < s$ and $a_{l} \neq x_{i}^{\ast}$, an incoming edge from $j_{s}$, and a threshold equal to the degree of $u_{i,j_{s}}$. Each node $i$ has threshold $1$ in the graph.
\end{lemma}

\begin{IEEEproof}
The nested canalyzing dynamics are equivalent to the condition $$\bigvee_{j_{s} \in \Omega_{i}}{\left(\left(\bigwedge_{j_{r} \in \Omega_{i}^{c} \cap \{1,\ldots,s\}}(\neg x_{a_{r}})\right) \wedge x_{j_{s}}\right)}.$$ Applying the construction of Section \ref{subsec:TSS_mapping} yields the graph described in the statement of the lemma.
\end{IEEEproof}

The following corollary provides a condition that admits computationally tractable approximation algorithms.

\begin{corollary}
\label{corollary:NC}
Let $s_{i}^{\ast} = \min{\{s : a_{i,s} = x_{i}^{\ast}\}}$. Consider an instance of the TSS problem defined by a graph $\hat{G} = (V,\hat{E})$, in which there is an edge $(j_{s},i) \in \hat{E}$ if $s \leq s_{i}^{\ast}$, where each node's threshold is equal to the degree of the node. Then a solution to this instance of TSS is sufficient to ensure convergence to a desired attractor.
\end{corollary}


The instance of TSS defined by Corollary \ref{corollary:NC} has a desirable structure, namely each node has a threshold equal to its degree (a unanimous threshold), equivalent to a Boolean AND decision rule. In undirected graphs, it is known that this Boolean decision rule is equivalent to the vertex cover problem \cite{chen2009approximability}. The following gives a necessary and sufficient condition for directed graphs 

\begin{proposition}
\label{prop:unanimous_threshold}
The condition of Corollary \ref{corollary:NC} holds if and only if each cycle in $\hat{G}$ contains at least one node from $S$ and each node is connected to at least one node in $S$.
\end{proposition}

\begin{IEEEproof}
First, suppose that a set $S$ does not satisfy the conditions of Corollary \ref{corollary:NC}, and yet the two conditions of the proposition hold. Let $i$ be a node satisfying $X_{i}^{\ast} = 0$. Then either $i$ is an isolated node, contradicting the assumption that all isolated nodes are in $S$, or there exists a neighbor, denoted $i_{1}$, satisfying $X_{i_{1}}^{\ast} = 0$. Proceeding inductively, we obtain a set of nodes $i_{0},i_{1},\ldots,i_{r}$ that all satisfy $X_{i_{j}}^{\ast} = 0$, and must either contain a cycle or have $X_{i_{r}}^{\ast}$ isolated, contradicting the conditions of the proposition.

Clearly if there is an isolated node $i \notin S$, then $X_{i}[k] \equiv 0$ for all $k$. Similarly, suppose that there is a set of edges $(i_{0},i_{1}),\ldots, (i_{m},i_{0})$ such that $S \cap \{i_{0},\ldots,i_{m}\} = \emptyset$. Then $X_{i_{l}}[0] = 0$,  and by induction $X_{i_{l}}[k] = 0$ for all $k$ since there exists a neighbor $i_{l-1}$ satisfying $X_{i_{l-1}}[k-1] = 0$.
\end{IEEEproof}


Note that this condition is the same as that of \cite{aswani2009graph}, but for the subgraph $\hat{G}$. Based on Proposition \ref{prop:unanimous_threshold}, we introduce an algorithm for selecting input genes under nested canalyzing dynamics as Algorithm \ref{algo:NC_dynamics}.

  \begin{center}
\begin{algorithm}[!htp]
	\caption{Algorithm for selecting a minimum-size set of genes to control a network with nested canalyzing dynamics.}
	\label{algo:NC_dynamics}
	\begin{algorithmic}[1]
		\Procedure{NC\_Dynamics}{$\hat{G}=(V,\hat{E})$}
            \State \textbf{Input}: Graph topology $\hat{G}=(V,\hat{E})$ constructed as in Corollary \ref{corollary:NC}.
            \State \textbf{Output}: Approximation of minimum-size input set $S$
            \State $S \leftarrow \emptyset$
            \State $\overline{G}=(\overline{V},\overline{E}) \leftarrow$ directed acyclic contraction of $\hat{G}$.
            \State $S \leftarrow S \cup \{v : v \mbox{ is an isolated singleton node of } \overline{G}\}$
            \For{$i \in \overline{V}$}
            \State $S_{i} \leftarrow FVS([i], G([i]))$ //FVS is algorithm of \cite{even1995approximating}
                        \State $S \leftarrow S \cup S_{i}$
            \EndFor
            \State \Return{$S$}
           	\EndProcedure
	 \end{algorithmic}
\end{algorithm}
\end{center}

Intuitively, Algorithm \ref{algo:NC_dynamics} is as follows. We first compute the maximal strongly connected subgraphs of $G$, which can be done in polynomial time, and contract with respect to these components to obtain a directed acyclic graph $\overline{G}$. It then suffices to ensure that each subgraph $G_{i}$ has no cycles that are disjoint from $S$, as well as ensuring that all nodes are connected to a node in $S$. This condition is ensured if each component is cycle-free and if all singleton isolated components (which are exactly the isolated nodes of $\hat{G}$) are in $S$. The  optimality guarantees of this approach are given in Proposition \ref{prop:NC_optimality}.

\begin{proposition}
\label{prop:NC_optimality}
Let $S^{\ast}$ denote the optimal set for the sufficient condition of Corollary \ref{corollary:NC}, and let $S$ denote the set returned by Algorithm \ref{algo:NC_dynamics}. Then $|S| \leq (\log{n})^{2}|S^{\ast}|$.
\end{proposition}

\begin{IEEEproof}
Define $S_{i}^{\ast} = S^{\ast} \cap [i]$, and $S_{i} = S \cap [i]$, so that $S^{\ast} = S_{1}^{\ast} \cup \cdots \cup S_{m}^{\ast}$ and $S = S_{1} \cup \cdots \cup S_{m}$. By \cite{even1995approximating}, $|S_{i}| \leq (\log{n})^{2} |S_{i}^{\ast}|$, and hence $$\frac{|S|}{|S^{\ast}|} \leq \sum_{i=1}^{m}{\frac{|S_{i}|}{|S_{i}^{\ast}|}} \leq (\log{n})^{2},$$ as desired.
\end{IEEEproof}

\begin{table*}
\centering
\caption{Number of Inputs for Control of Biological Networks}
\label{table:real-world-sim}
\begin{tabular}{|c|c|c|}
\hline
\textbf{Network} & \textbf{Number of Nodes (Edges)} & \textbf{Number of Inputs} \\
\hline
Apoptosis & 39 (70) & 10 \\
\hline
\emph{Bordatella Bronchiseptica} & 33 (79) & 2 \\
\hline
Breast Cell Development Network & 21 (70) & 7 \\
\hline
Mammalian Cell Cycle & 19 (48) & 3\\
\hline
T-Cell Differentiation & 19 (30) & 10 \\
\hline
T-Cell Signaling &  37 (48) & 3 \\
\hline
\end{tabular}
\end{table*}

\section{Generalizations to Probabilistic Graphs and Cyclic Attractors}
\label{sec:probabilistic}
In this section, we investigate two generalizations to the problem formulation. We first investigate probabilistic Boolean networks, followed by guaranteeing convergence to attractors with multiple states.

\subsection{Probabilistic Regulatory Networks}
\label{subsec:probabilistic}
Probabilistic Boolean networks are an extension of Boolean regulatory networks to model the inherent uncertainty of biological systems. A Boolean regulatory network is defined by a graph $G=(V,E)$ and a set of $K$ update functions $f(\cdot, 1), \ldots, f(\cdot, K)$ each of which maps $2^{|V|}$ into $2^{|V|}$. The Boolean network is also characterized by a random process $\xi(t) \in \{1,\ldots,K\}$, so that $\mathbf{x}(k+1) = f(\mathbf{x}(k), \xi(t))$. 

A generalization of the approach of Section \ref{subsec:TSS_mapping} is as follows. Let $$f_{i}(\mathbf{x},j) = (y_{11}^{(j)} \vee \cdots \vee y_{1r_{1}}^{(j)}) \wedge \cdots \wedge (y_{l1}^{(j)} \vee \cdots \vee y_{lr_{1}}^{(j)})$$ be a CNF realization of the dynamics of node $i$ in topology $j$. Furthermore, define an extended function $f$ by 
\begin{equation}
\overline{f}_{i,1}(\mathbf{x}) = \left(\bigvee_{j=1}{f_{i}(\mathbf{x},j)}\right)  
\wedge \left(\bigwedge_{j=1}^{n}{f_{i}(\mathbf{x}_{-i}, x_{i}^{\ast},j)}\right),
\end{equation}
where $f_{i}(\mathbf{x}_{-i},x_{i}^{\ast},j)$ refers to the value of $f_{i}(\cdot, j)$ when $x_{i} = x_{i}^{\ast}$ and all other indices are equal to $\mathbf{x}$. Define $\overline{f}_{i,0}$ by 
\begin{displaymath}
\overline{f}_{i,0}(\mathbf{x}) = \left(\bigwedge_{j=1}^{n}{f_{i}(\mathbf{x},j)}\right) 
\vee \left(\bigvee_{j=1}^{n}{f_{i}(\mathbf{x}_{-i}, x_{i}^{\ast},j)}\right),
\end{displaymath}
The definition of $f_{i,1}$ (resp. $f_{i,0}$) is chosen so that, if $\overline{f}_{i,1}(\mathbf{x}^{\ast}) = 1$, then $f_{i}(\mathbf{x},j) = 1$ for at least one function $j$. Furthermore, if node $i$ achieves the desired attractor, then the function will remain in the desired state.

\begin{proposition}
\label{prop:probabilistic_sufficient}
Let $G=(V,E)$ be a Boolean network with update functions $f(\cdot,1),\ldots,f(\cdot,K)$. Construct an instance of TSS based on the approach of Section \ref{subsec:TSS_mapping}, using the functions $\overline{f}_{i,0}$ and $\overline{f}_{i,1}$. Then the resulting set $S$ is sufficient to guarantee convergence to an attractor $\mathbf{x}^{\ast}$, provided that $f(\mathbf{x}^{\ast},j) = \mathbf{x}^{\ast}$ for all $j$.
\end{proposition}

\begin{IEEEproof}
The approach is to show by induction that, if $i \in X[k]$, then $x_{i}(t)$ eventually reaches $x_{i}^{\ast}$ regardless of the set of topologies. Suppose the result is true up to iteration $k$. By construction of $\overline{f}_{i,0}$, there exists at least one time step $T$ such that $x_{i}(T) = x_{i}^{\ast}$, and by construction of $\overline{f}_{i,1}$, $x_{i}(t) = x_{i}^{\ast}$ for all $t \geq T$.
\end{IEEEproof}

In the special case of threshold dynamics, we have the following corollary.

\begin{corollary}
\label{corollary:probabilistic_threshold}
For networks with threshold dynamics, if a node is excitatory or inhibitory under all functions $f^{(i)}$ and only the threshold varies, then the threshold $\overline{\tau}_{i} = \max{\{\tau_{i}^{(1)},\ldots,\tau_{i}^{(K)}\}}$ is sufficient to ensure convergence.
\end{corollary}


Another relevant class of models arises when different nodes may update their states asynchronously. The resulting network has $K=n$, where $n$ is the number of nodes. The functions $f(\cdot, i)$ are defined by 
\begin{displaymath}
f_{j}(\mathbf{x}, i) = \left\{
\begin{array}{ll}
f_{i}(\mathbf{x}), & j = i \\
x_{j}, & j \neq i
\end{array}
\right.
\end{displaymath}
so that only node $i$ updates its state and all other nodes maintain fixed state values. 

\begin{lemma}
\label{lemma:async_update}
Suppose that the set $S$ satisfies the conditions of Proposition \ref{prop:TSS_sufficient}. Then the set $S$ is sufficient to guarantee convergence to a desired attractor under asynchronous dynamics.
\end{lemma}

\begin{IEEEproof}
Let $i_{1}, i_{2}, \ldots, i_{n}$ denote the sequence in which nodes are activated by the process $X[k]$. Define $T_{1}$ as $T_{1} = \min{\{t: \xi(t) = i_{1}\}}$ and $T_{j}$ for $j \geq 2$ as $$T_{j} = \min{\{t: \xi(t) = i_{j}, t > T_{j-1}\}}.$$ 

We have that $x_{i_{1}}(T_{1}) = x_{i_{1}}^{\ast}$. Proceeding inductively, at time $T_{j}$, a set of neighbors of $N_{i_{j}}$ has reached the desired attractor, which is sufficient to ensure that $X_{i_{j}}[k] = 1$. Hence, $x_{i_{j}}(t) = x_{i_{j}}^{\ast}$ for $t \geq T_{j}$.
\end{IEEEproof}

\subsection{Convergence to Cyclic Attractors}
\label{subsec:cyclic}
We now remark on the gene selection problem to ensure convergence to a \emph{cyclic attractor}, i.e., an attractor that consists of multiple states. We formulate a TSS-based condition that is analogous to the fixed point condition of Section \ref{subsec:TSS_mapping}. Let $(\mathbf{x}^{1}, \ldots, \mathbf{x}^{p})$ denote the desired attractor. 

The approach is to construct a network graph $\hat{G} = (\hat{V}, \hat{E})$, where $\hat{V} = \hat{V}_{1} \cup \cdots \cup \hat{V}_{p}$. Using the CNF form, we have $\hat{V}_{a} = \{i_{a} : i \in V\} \cup \{j_{i_{s},a} : s =1,\ldots,l\}$ for $a=1,\ldots,p$. The edge set is defined by 
\begin{IEEEeqnarray*}{rCl}
\hat{E} &=& \{(j_{i_{s},a},i_{a}) : s=1,\ldots,l, a=1,\ldots,p\} \\
&& \cup \{(j_{a}, b_{i_{s},(a+1)}) : x_{j} \in \left\{y_{su} : u=1,\ldots,l_{s}\},\right.\\
&& \quad \left. x_{j}^{a\ast} = x_{i}^{(a+1)\ast},a=1,\ldots,(p-1)\right\} \\
&& \cup \left\{(j_{a}, b_{i_{s},(a+1)}) : \neg x_{j} \in \{y_{su} : u=1,\ldots,l_{s}\}, \right. \\
&& \quad \left. x_{j}^{a\ast} \neq x_{i}^{(a+1)\ast},a=1,\ldots,(p-1)\right\} \\
&& \cup \{(j_{p},b_{i_{s},1}) : x_{j} \in \{y_{su} : u=1,\ldots,l_{s}\}, x_{j}^{p\ast} = x_{i}^{1\ast}\} \\
&& \cup \{(j_{p}, b_{i_{s},1}) : \neg x_{j} \in \{y_{su} : u=1,\ldots,l_{s}\}, x_{j}^{p\ast} \neq x_{i}^{1\ast}\} \\
\end{IEEEeqnarray*}

This definition is analogous to Section \ref{subsec:TSS_mapping}, except there is an edge from the $a$-th copy of $V$ to the $(a+1)$-th copy if the state of node $j$ in the $a$-th state of the attractor influences the state of node $i$ in the $(a+1)$-th state of the attractor. The thresholds are defined as in Section \ref{subsec:TSS_mapping}.

\begin{proposition}
\label{prop:cyclic-attractor}
If $S$ is a target set for the graph $\hat{G}$ with thresholds $\tau$, then controlling the set of genes $S$ ensures convergence to the desired cyclic attractor.
\end{proposition}

The proof is  omitted due to space constraints.


\begin{figure*}
\centering
$\begin{array}{ccc}
\includegraphics[width=2in]{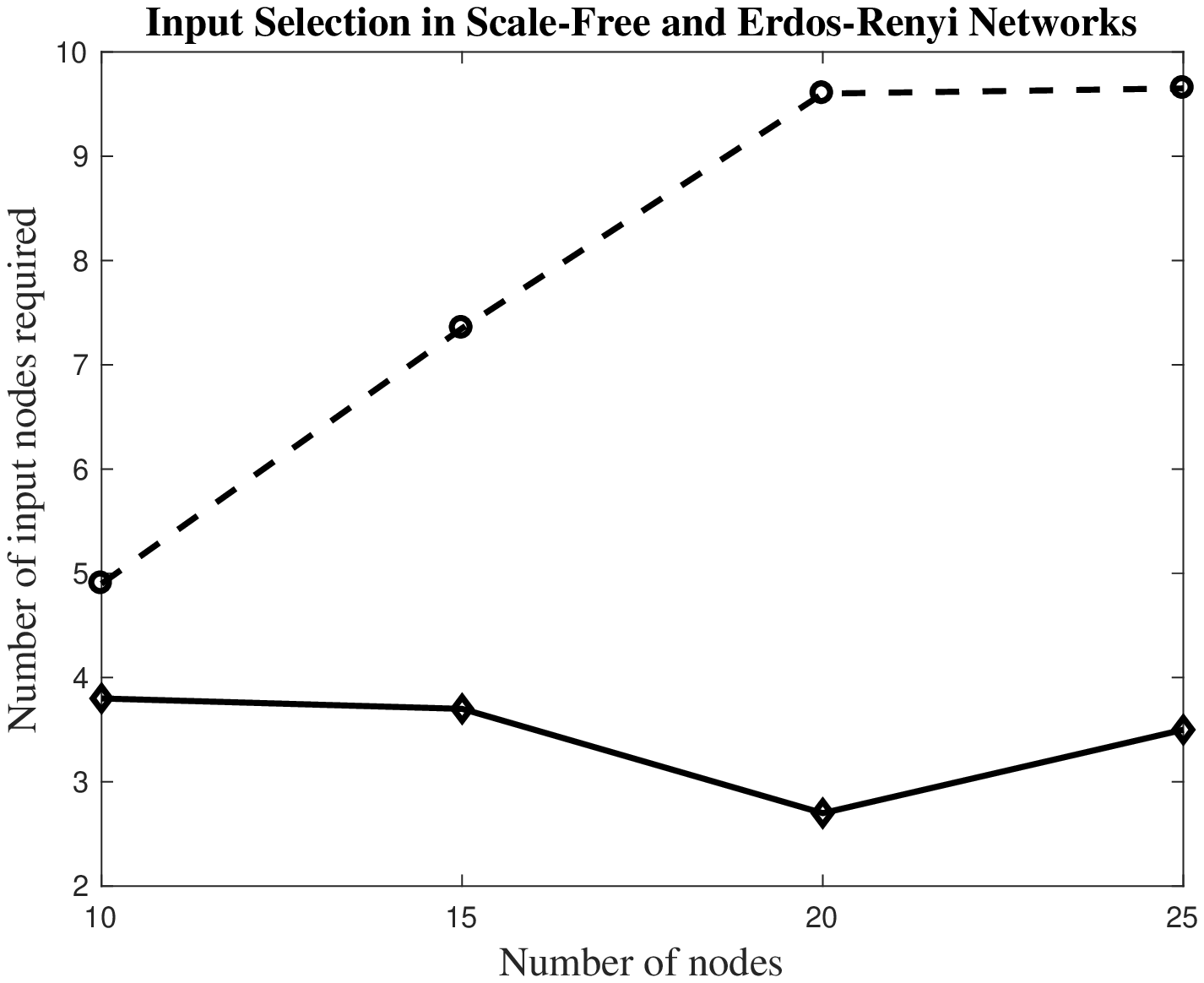} & 
\includegraphics[width=2in]{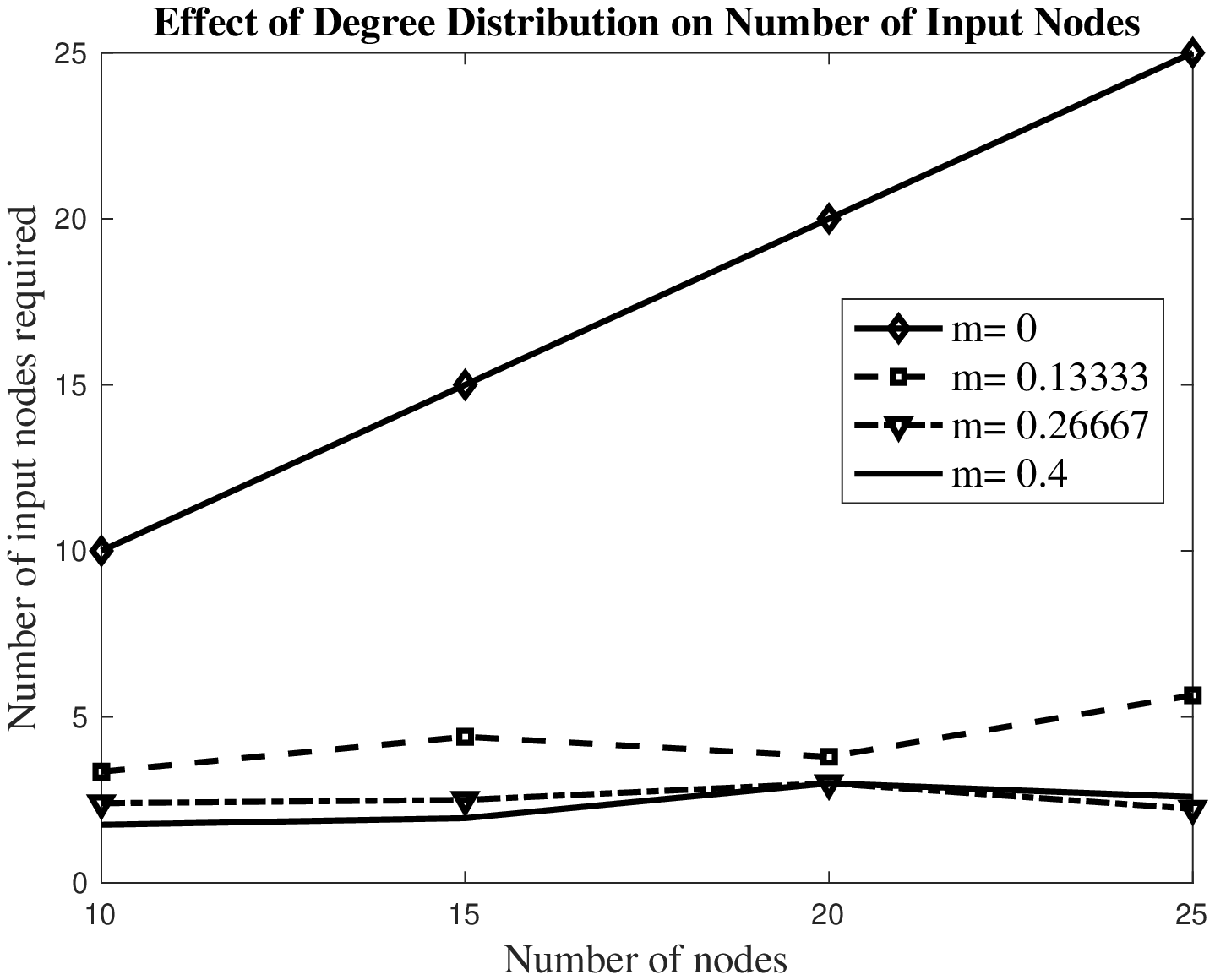} &
\includegraphics[width=2in]{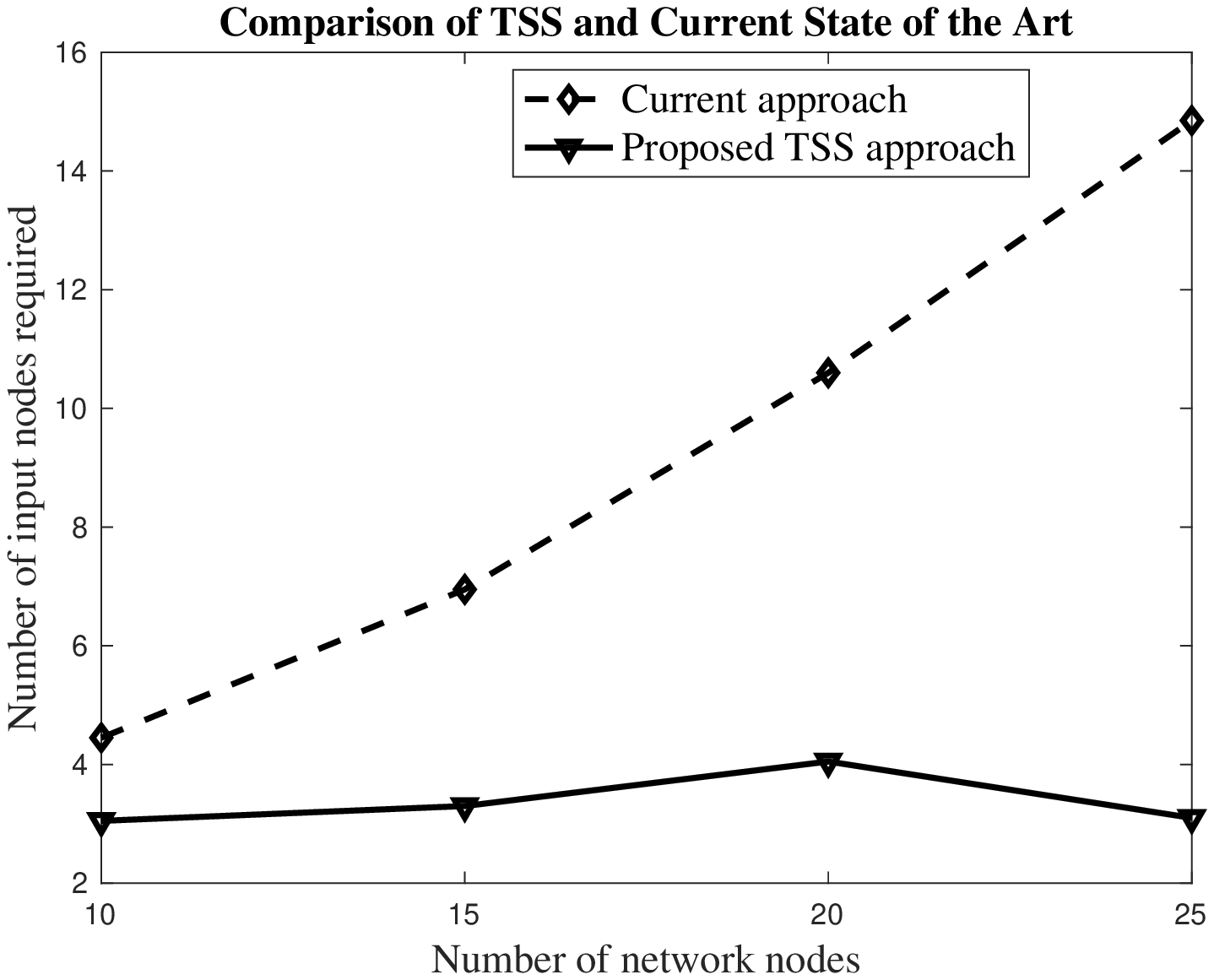} \\
\mbox{(a)} & \mbox{(b)} & \mbox{(c)}
\end{array}$
\caption{Empirical results on input selection for randomly generated regulatory networks. (a) Comparison of number of input nodes needed for Erdos-Renyi and scale-free networks. The scale-free network requires fewer input nodes even when parameters are chosen to maintain the same average node degree, due to the increased  clustering and presence of high-degree hubs in such networks. (b) Effect of degree distribution on number of input nodes required. Increasing the parameter $m$ results in a higher node degree. Thus, networks with large node degree require fewer inputs. (c) Comparison between our proposed TSS and the current state of the art. The TSS approach consistently requires fewer inputs.}
\label{fig:simulation}
\end{figure*}

\section{Numerical Study}
\label{sec:simulation}
We conducted a numerical study of our approach using Matlab$^{TM}$. The goals of our numerical study were two-fold. The first objective was to evaluate the behavior of our approach on a real-world biological network. The second objective was to observe trends in the number of input nodes required to converge to a desired attractor, as a function of parameters such as the class of network (e.g., scale-free or Erdos-Renyi graph), the average node degree,  and the number of nodes in the network. For all threshold networks, a fixed-point attractor was computed by solving an integer linear program.

In order to complete the first objective, we obtained several biological regulatory networks from the Cell Collective website~\cite{cellcollective}, which maintains an archive of biological networks including the topology and the Boolean dynamics of each node. We evaluated our approach on several networks, including the T-cell differentiation network, the yeast cell cycle network, an Apoptosis network, a model of the mammalian cell cycle, the T-cell signaling network, and the regulatory network of \emph{Bordatella Bronchiseptica}. For the node dynamics, we used threshold dynamics with threshold $0$, in which nodes were chosen as excitatory or inhibitory based on the published Boolean dynamics. The results are summarized in Table \ref{table:real-world-sim}.

From Table \ref{table:real-world-sim}, we observe that the number of inputs required is typically a small fraction of the number of network nodes, implying that only a few inputs are needed to guarantee convergence to a desired attractor. The exception to this rule is the T-Cell differentiation network, in which nearly half of the nodes must act as inputs.

We then evaluated the behavior of our algorithms on synthetic networks with different topologies. We first performed a comparison of classical random graph models with graph models that more closely approximate regulatory networks. We chose Erdos-Renyi and scale-free graphs for comparison. In an Erdos-Renyi graph, each node is connected to each other node with a fixed probability $p$. In a scale-free graph, each node is connected to $m$ randomly chosen nodes, where the probability of an edge is proportional to the degree of the node (preferential attachment model). As shown in Figure \ref{fig:simulation}(a), we found that scale-free networks consistently require fewer input nodes than Erdos-Renyi random graphs to guarantee convergence. This may be due to the presence of high-degree hubs and increased clustering in scale-free networks. Indeed, the number of input nodes required by the scale-free network did not increase as a function of the network size.

We studied the effect of the degree distribution on the number of input nodes required. We considered scale-free networks in which the degree distribution is varied by changing $m$. For all cases we considered threshold dynamics with threshold $0$ and edges randomly assigned as excitatory or inhibitory with probability $0.5$. We found that high-degree networks required fewer inputs, as a subset of well-connected hub nodes are sufficient to guarantee convergence (Figure \ref{fig:simulation}(b)).

Finally, we compared our approach to a current state of the art approach (Figure \ref{fig:simulation}(c)), which is based on selecting a minimum-size set of inputs such that all cycles contain at least one input \cite{aswani2009graph}. The network considered was a scale-free graph with $m=0.2n$, where $n$ is the number of nodes. The cycle-based method consistently required more input nodes than our proposed TSS-based algorithm. This result agrees with the theoretical guarantees of Proposition \ref{prop:weaker}.

\section{Conclusions and Future Work}
\label{sec:conclusion}
This paper investigated the problem of selecting input nodes to control biological regulatory networks. Under a Boolean network model, we formulated the problem of selecting a minimum-size set of inputs to guarantee convergence to a desired attractor, defined as a stable fixed point of the network dynamics, and showed that this problem cannot be approximated up to any provable bound unless P=NP. We showed that a sufficient condition for convergence can be mapped to an instance of the target set selection problem, which is defined as selecting a minimum-size set of nodes to ensure that all nodes are activated by a threshold dynamics. 

We analyzed our sufficient condition under biologically relevant special cases of the network dynamics. For threshold dynamics with modular structure, we proposed polynomial-time exact algorithms for input selection. In networks with hierarchical structure, we introduced an $O(n^{2})$ algorithm that selects a minimum-size input set up to a provable bound of $\log{n}$. Finally, in networks with nested canalyzing dynamics, we showed that a sufficient condition for convergence to a desired attractor is ensuring that each cycle in a subgraph contains at least one input node, leading to polynomial-time algorithms with an optimality bound of 2. We proposed generalizations of our approach to asynchronous and probabilistic dynamics, as well as multi-state attractors.

We plan to investigate tighter sufficient conditions and  exploit additional network structures to improve computation times. Furthermore, other control actions, such as time-varying interventions and changes in the network topology, will be considered in future. Finally, computing the number of distinct minimum-size input sets is an additional open research problem. 


\bibliographystyle{IEEEtran}
\bibliography{TCNS_2017}

\begin{thebibliography}{10}
\providecommand{\url}[1]{#1}
\csname url@samestyle\endcsname
\providecommand{\newblock}{\relax}
\providecommand{\bibinfo}[2]{#2}
\providecommand{\BIBentrySTDinterwordspacing}{\spaceskip=0pt\relax}
\providecommand{\BIBentryALTinterwordstretchfactor}{4}
\providecommand{\BIBentryALTinterwordspacing}{\spaceskip=\fontdimen2\font plus
\BIBentryALTinterwordstretchfactor\fontdimen3\font minus
  \fontdimen4\font\relax}
\providecommand{\BIBforeignlanguage}[2]{{%
\expandafter\ifx\csname l@#1\endcsname\relax
\typeout{** WARNING: IEEEtran.bst: No hyphenation pattern has been}%
\typeout{** loaded for the language `#1'. Using the pattern for}%
\typeout{** the default language instead.}%
\else
\language=\csname l@#1\endcsname
\fi
#2}}
\providecommand{\BIBdecl}{\relax}
\BIBdecl

\bibitem{alon2006introduction}
U.~Alon, \emph{{An Introduction to Systems Biology: Design Principles of
  Biological Circuits}}.\hskip 1em plus 0.5em minus 0.4em\relax CRC press,
  2006.

\bibitem{kauffman1993origins}
S.~A. Kauffman, \emph{{The Origins of Order: Self Organization and Selection in
  Evolution}}.\hskip 1em plus 0.5em minus 0.4em\relax Oxford University Press,
  USA, 1993.

\bibitem{huang2005cell}
S.~Huang, G.~Eichler, Y.~Bar-Yam, and D.~E. Ingber, ``Cell fates as
  high-dimensional attractor states of a complex gene regulatory network,''
  \emph{{Physical Review Letters}}, vol.~94, no.~12, p. 128701, 2005.

\bibitem{huang2009cancer}
S.~Huang, I.~Ernberg, and S.~Kauffman, ``Cancer attractors: a systems view of
  tumors from a gene network dynamics and developmental perspective,'' in
  \emph{{Seminars in Cell \& Developmental Biology}}, vol.~20, no.~7.\hskip 1em
  plus 0.5em minus 0.4em\relax Elsevier, 2009, pp. 869--876.

\bibitem{davidich2008boolean}
M.~I. Davidich and S.~Bornholdt, ``Boolean network model predicts cell cycle
  sequence of fission yeast,'' \emph{PloS One}, vol.~3, no.~2, p. e1672, 2008.

\bibitem{li2004yeast}
F.~Li, T.~Long, Y.~Lu, Q.~Ouyang, and C.~Tang, ``The yeast cell-cycle network
  is robustly designed,'' \emph{Proceedings of the National Academy of Sciences
  of the United States of America}, vol. 101, no.~14, pp. 4781--4786, 2004.

\bibitem{shmulevich2010probabilistic}
I.~Shmulevich and E.~R. Dougherty, \emph{{Probabilistic Boolean Networks: The
  Modeling and Control of Gene Regulatory Networks}}.\hskip 1em plus 0.5em
  minus 0.4em\relax SIAM, 2010.

\bibitem{macarthur2009systems}
B.~D. MacArthur, A.~Ma'ayan, and I.~R. Lemischka, ``Systems biology of stem
  cell fate and cellular reprogramming,'' \emph{Nature Reviews Molecular Cell
  Biology}, vol.~10, no.~10, pp. 672--681, 2009.

\bibitem{wu2015network}
L.~Wu, Y.~Shen, M.~Li, and F.-X. Wu, ``Network output controllability-based
  method for drug target identification,'' \emph{IEEE Transactions on
  Nanobioscience}, vol.~14, no.~2, pp. 184--191, 2015.

\bibitem{kim2013discovery}
J.~Kim, S.-M. Park, and K.-H. Cho, ``Discovery of a kernel for controlling
  biomolecular regulatory networks,'' \emph{Scientific Reports}, vol.~3, 2013.

\bibitem{ackerman2010combinatorial}
E.~Ackerman, O.~Ben-Zwi, and G.~Wolfovitz, ``Combinatorial model and bounds for
  target set selection,'' \emph{Theoretical Computer Science}, vol. 411,
  no.~44, pp. 4017--4022, 2010.

\bibitem{lahdesmaki2003learning}
H.~L{\"a}hdesm{\"a}ki, I.~Shmulevich, and O.~Yli-Harja, ``On learning gene
  regulatory networks under the {B}oolean network model,'' \emph{Machine
  Learning}, vol.~52, no. 1-2, pp. 147--167, 2003.

\bibitem{chaves2005robustness}
M.~Chaves, R.~Albert, and E.~D. Sontag, ``Robustness and fragility of {B}oolean
  models for genetic regulatory networks,'' \emph{Journal of Theoretical
  Biology}, vol. 235, no.~3, pp. 431--449, 2005.

\bibitem{karlebach2008modelling}
G.~Karlebach and R.~Shamir, ``Modeling and analysis of gene regulatory
  networks,'' \emph{Nature Reviews Molecular Cell Biology}, vol.~9, no.~10, pp.
  770--780, 2008.

\bibitem{bornholdt2008boolean}
S.~Bornholdt, ``Boolean network models of cellular regulation: prospects and
  limitations,'' \emph{Journal of the Royal Society Interface}, vol.~5, no.
  Suppl 1, pp. S85--S94, 2008.

\bibitem{waddington1940organisers}
C.~H. Waddington \emph{et~al.}, \emph{{Organisers and Genes}}.\hskip 1em plus
  0.5em minus 0.4em\relax Cambridge Biological Studies, 1940.

\bibitem{aswani2009graph}
A.~Aswani, N.~Boyd, and C.~Tomlin, ``Graph-theoretic topological control of
  biological genetic networks,'' in \emph{2009 American Control
  Conference}.\hskip 1em plus 0.5em minus 0.4em\relax IEEE, 2009, pp.
  1700--1705.

\bibitem{kearney2016framework}
G.~Kearney and M.~Fardad, ``On a framework for analysis and design of cascades
  on boolean networks,'' \emph{55th IEEE Conference on Decision and Control
  (CDC)}, pp. 997--1002, 2016.

\bibitem{liu2011controllability}
Y.-Y. Liu, J.-J. Slotine, and A.-L. Barab{\'a}si, ``Controllability of complex
  networks,'' \emph{Nature}, vol. 473, no. 7346, pp. 167--173, 2011.

\bibitem{ben2011treewidth}
O.~Ben-Zwi, D.~Hermelin, D.~Lokshtanov, and I.~Newman, ``Treewidth governs the
  complexity of target set selection,'' \emph{Discrete Optimization}, vol.~8,
  no.~1, pp. 87--96, 2011.

\bibitem{chiang2013some}
C.-Y. Chiang, L.-H. Huang, B.-J. Li, J.~Wu, and H.-G. Yeh, ``Some results on
  the target set selection problem,'' \emph{Journal of Combinatorial
  Optimization}, vol.~25, no.~4, pp. 702--715, 2013.

\bibitem{herrero2001hierarchical}
J.~Herrero, A.~Valencia, and J.~Dopazo, ``A hierarchical unsupervised growing
  neural network for clustering gene expression patterns,''
  \emph{Bioinformatics}, vol.~17, no.~2, pp. 126--136, 2001.

\bibitem{ravasz2002hierarchical}
E.~Ravasz, A.~L. Somera, D.~A. Mongru, Z.~N. Oltvai, and A.-L. Barab{\'a}si,
  ``Hierarchical organization of modularity in metabolic networks,''
  \emph{Science}, vol. 297, no. 5586, pp. 1551--1555, 2002.

\bibitem{chen2009approximability}
N.~Chen, ``On the approximability of influence in social networks,'' \emph{SIAM
  Journal on Discrete Mathematics}, vol.~23, no.~3, pp. 1400--1415, 2009.

\bibitem{even1995approximating}
G.~Even, J.~S. Naor, B.~Schieber, and M.~Sudan, ``Approximating minimum
  feedback sets and multi-cuts in directed graphs,'' in \emph{International
  Conference on Integer Programming and Combinatorial Optimization}.\hskip 1em
  plus 0.5em minus 0.4em\relax Springer, 1995, pp. 14--28.

\bibitem{cellcollective}
``{The Cell Collective},'' http://www.cellcollective.org.

\end{thebibliography}

%
\IEEEpeerreviewmaketitle

\end{document}